\documentclass[twocolumn]{emulateapj}
\usepackage{graphicx}

\def\gr{$\gamma$-ray}

\shortauthors{Neronov et al.}
\shorttitle{Degree-scale "jets" from TeV blazars}

\begin{document}
\author{A.~Neronov$^{1}$, D.~Semikoz$^{2,3}$, M.~Kachelriess$^{4}$, S.~Ostapchenko$^{4,5}$, A.~Elyiv$^{6,7}$}
\affil{$^{1}$ ISDC Data Centre for Astrophysics, Ch. d'Ecogia 16, Versoix, Switzerland\\
$^{2}$ APC, 10 rue Alice Domon et Leonie Duquet, F-75205 Paris Cedex 13, France\\
$^{3}$ Institute for Nuclear Research RAS, 60th October Anniversary prosp. 7a, Moscow, 117312, Russia\\
$^{4}$Institutt for fysikk, NTNU, Trondheim, Norway\\
$^{5}$D. V. Skobeltsyn Institute of Nuclear Physics, Moscow State University, Russia\\
$^{6}$ Institut d'Astrophysique et de Geophysique, Universite de Liege, 4000 Liege,
Belgium\\
$^{7}$ Main Astronomical Observatory, Academy of Sciences of Ukraine, 27 Akademika
Zabolotnoho St., 03680 Kyiv, Ukraine}

\title{Degree-scale GeV ``jets" from active and dead TeV blazars}

\begin{abstract}
We show that images of TeV blazars in the GeV energy band should contain, along 
with point-like sources,  degree-scale jet-like extensions. These GeV extensions 
are the result of electromagnetic cascades initiated by TeV \gr s 
interacting with extragalactic background light and the deflection of the 
cascade electrons/positrons in extragalactic magnetic fields (EGMF).  Using 
Monte-Carlo simulations, we study the spectral and timing properties of the 
degree-scale extensions in simulated GeV band images of TeV blazars. 
We show that the brightness profile of such degree-scale extensions can be used to 
infer the lightcurve of the primary TeV \gr\ source over the past $10^7$~yr, i.e.\ 
over a time scale comparable to the life-time of the parent active galactic nucleus. 
This implies that the degree-scale jet-like GeV emission could be detected not only 
near known active TeV blazars, but also from ``TeV blazar remnants", whose central 
engines were switched off up to ten million years ago. Since the brightness profile of 
the GeV ``jets'' depends on the strength and the structure of the EGMF, their 
observation provides additionally information about the EGMF.
\end{abstract}


\maketitle

\textit{Introduction.}
Significant progress in understanding the activity of blazars, i.e.\
active galaxies with relativistic jets aligned with the line of sight, was achieved 
with the start of operation of the {\it Fermi\/} telescope. The combination of data 
from {\it Fermi\/} in the 0.1 -- 10~GeV energy band and from ground based \gr\ 
telescopes like HESS, MAGIC and VERITAS in the 100~GeV -- 10~TeV band provides 
detailed simultaneous spectral and timing information for the most extreme 
representatives of the blazar population \citep{abdo09}.

The TeV \gr\ flux from distant blazars is significantly attenuated 
by pair production on the infrared/optical extragalactic 
background light (EBL) \citep{kneiske04,Stecker06,franceschini08,Primack08}. TeV \gr s
that are absorbed on the way from the primary \gr\ source initiate electromagnetic 
cascades in the intergalactic space. The charged component of the electromagnetic cascade is deflected by the 
EGMF. Potentially observable effects of such electromagnetic 
cascades in the EGMF include the 
``echoes" of multi-TeV \gr\ 
flares~\citep{plaga,japanese} and the appearance of extended emission around initially 
point-like \gr\ sources \citep{coppi,neronov07,kachelriess09,elyiv09}.

\begin{figure}
\includegraphics[width=\linewidth]{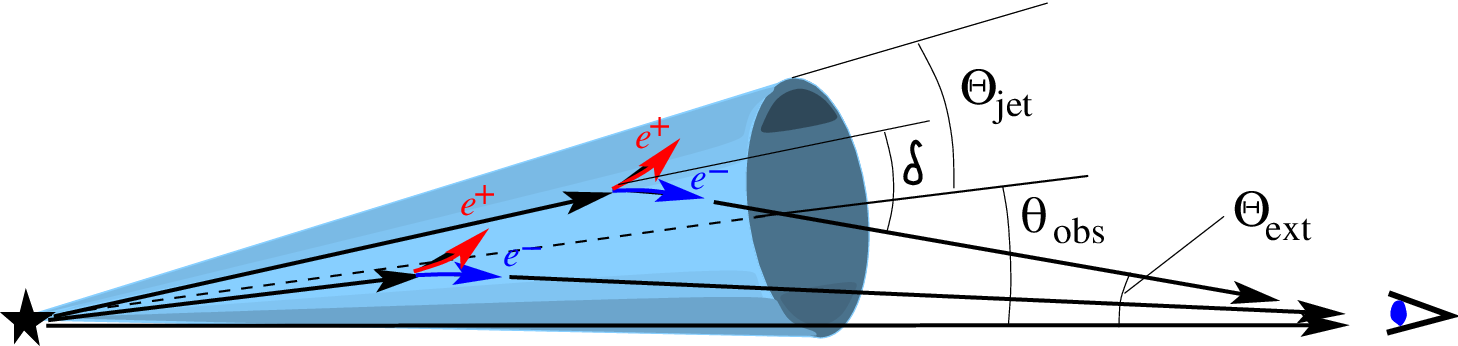}
\caption{Geometry of the propagation of direct and cascade \gr s from the source 
(on the left) to the observer (on the right).}
\label{fig:scheme}
\end{figure}

TeV \gr\ emission from blazars is believed to be relativistically beamed into a narrow 
cone (jet) with an opening angle 
$\Theta_{\rm jet}\sim \Gamma^{-1}\sim 5^\circ\left[\Gamma/10\right]$, where $\Gamma$ is the 
bulk Lorentz factor of the \gr\ emitting plasma. Blazars are a special type of \gr\ 
emitting AGN for which the angle between the line of sight (LOS) and the jet axis, 
$\theta_{\rm obs}$, is $\theta_{\rm obs}\lesssim\Theta_{\rm jet}$, see Fig.~\ref{fig:scheme} 
\citep{urry}. 

In general, the number of blazars with a given jet-LOS misalignment angle is expected 
to scale as $dN/d\theta_{\rm obs}\sim \theta_{\rm obs}$ in the range 
$0<\theta_{\rm obs}\le \Theta_{\rm jet}$. Thus most observed TeV blazars 
should have $\theta_{\rm obs}\sim \Theta_{\rm jet}$, rather than $\theta_{\rm obs}\simeq 0$. 
Consequently, the TeV \gr\ emission pattern is not symmetric with respect to the 
axis source-observer. In sources with $\theta_{\rm obs}\sim \Theta_{\rm jet}$, most 
multi-TeV \gr s are emitted preferentially on one side of the LOS, as it is shown in 
Fig. \ref{fig:scheme}. As a result, the extended emission from the cascade initiated 
by the absorption of TeV \gr s in interactions with EBL photons should appear as 
one- or two-sided jet-like extension next to the primary point source~\citep{aharonian_book}, rather than as an extended 
axially symmetric halo discussed previously~\citep{coppi,neronov07,neronov09,kachelriess09,elyiv09}. 

In what follows we discuss the spectral and timing properties of such jet-like 
cascade extensions in the 0.1-1~GeV images of TeV blazars. Our study is
based on two independent Monte Carlo codes for  \gr\ induced 
electromagnetic cascades in the intergalactic space, introduced  
by \cite{kachelriess09} and \cite{elyiv09}.

\textit{Basic formulae.}
Before presenting our numerical results, we discuss the basic physics of the
phenomenon in a simplified picture. In particular, we assume that the
electromagnetic cascade consists only of two steps and replace probability
distributions by their means.
Then the mean free path of VHE \gr s through the EBL can be approximated by 
$D_\gamma(E_{\gamma_0}) = \kappa \left[E_{\gamma_0}/1\mbox{ TeV}\right]^{-1}\mbox{ Gpc}$,
where the numerical factor $\kappa=\kappa(E_{\gamma_0},z)\sim {\cal O}(1)$ 
includes the uncertainties of the EBL modeling. Pair production on 
EBL reduces the flux of $\gamma $-rays from the source by a factor 
$\exp\left[-\tau(E_{\gamma_0})\right]$, where $\tau\simeq D/D_\gamma$ is the optical depth 
with respect to pair production and $D$ is the distance to the source. 

Electron-positron pairs created in interactions of multi-TeV $\gamma$-rays with EBL 
photons produce secondary $\gamma$-rays via inverse Compton (IC) scattering on 
Cosmic Microwave Background (CMB) photons. Typical energies of the IC photons reaching 
the Earth are $E_{\gamma} = (4/3)\epsilon_{\rm CMB}E_e^2/m_e^2\simeq 0.8
\left[E_{\gamma_0}/1 \mbox{ TeV}\right]^2\mbox{ GeV}$
where $\epsilon_{\rm CMB}=6\times 10^{-4}$~eV is the typical energy of CMB photons. 

Deflections of $e^+e^-$ pairs produced by the $\gamma$-rays, which were initially emitted 
away from the observer, can redirect secondary photons toward the observer. This effect 
leads to the appearance of extended emission around an initially point source of \gr s 
\citep{neronov07,kachelriess09,elyiv09}. 

In the absence of perfect alignment of the jet axis with the line of sight, the extended 
cascade emission might be strongly asymmetric. It might appear as a jet-like feature 
next to the primary \gr\ source. The maximal angular size of this jet-like feature can be 
estimated as the size of the projected \gr\ mean free path as
$\Theta_{\rm ext,max}\simeq D_\gamma\theta_{\rm obs}/(D-D_\gamma)$,
if $D_\gamma<D$ (i.e. $\tau>1$).  If $\tau(E_{\gamma_0})<1$, the cascade emission from the 
TeV \gr\ beam can extend to very large angles $\Theta_{\rm ext,max}\sim \pi/2$.

The jet-like extended emission can be observed only if deflections of the cascade $e^+e^-$
pairs are sufficiently large to redirect the cascade emission toward the observer. If the 
correlation length  of the EGMF is larger than the electron cooling distance
$D_e=3m_e^2c^3/(4\sigma_TU_{\rm CMB}E_e)\simeq  0.7\left[E_e/0.5\mbox{ TeV}\right]^{-1}\mbox{ Mpc}$,
where $\sigma_T$ denotes the Thomson cross section and $U_{\rm CMB}$ the energy density of the 
CMB photons, then the deflection angle can be estimated as  \citep{neronov09}
$\delta=D_e/R_L\simeq 3^\circ\left[B/10^{-17}\mbox{ G}\right]\left[E_e/0.5\mbox{ TeV}\right]^{-2}$
with $R_L$ as the Larmor radius of electrons and positrons. 

If the correlation length $\lambda_B$ of the EGMF is much smaller than the electron cooling 
distance $D_e$, the deflection angle can be  estimated using the diffusion approximation as
$\delta=\sqrt{D_e\lambda_B}/R_L\simeq 3^\circ\left[E_e/0.5\mbox{ TeV}\right]^{-3/2}\left[B/10^{-17}\mbox{ G}\right]\left[\lambda_B/0.7\mbox{ Mpc}\right]^{1/2}$.
If the EGMF is weak, electron/positron trajectories are not strongly deflected during one
cooling time and thus secondary cascade \gr s are emitted within a cone with opening 
angle of order ${\cal O}(\delta)$. In this case only a part of the cascade emission could be observed. If the mean free path of the primary \gr s is much shorter than the distance to the source, the angular extension could be estimated from the simple geometrical consideration of Fig. \ref{fig:scheme} as 
$\sin\left( \Theta_{\rm ext}(B)\right)= (D_\gamma/D)\sin\delta$.
Otherwise, the angular size of the source is found from the sum of the angles of triangle with vertices at the source, at the pair production point and at the position of the observer (see Fig. \ref{fig:scheme}) as 
$ \Theta_{\rm ext}(B)=\delta-\theta_{\rm obs}$.

Most of the known TeV blazars have moderate distances, so that 
$\tau(E_{\gamma_0}=1\mbox{ TeV})\le 1$. In this case, $\Theta_{\rm ext, max}\sim \pi/2$ and 
$\Theta_{\rm ext}(B)=\delta-\theta_{\rm obs}$.  A measurement of $\Theta_{\rm ext}\ll \Theta_{\rm ext, max}$ thus provides a measurement of $\delta$ and, in this way, gives a constraint on 
the parameters of the EGMF, i.e.\ $B$ and $\lambda_B$.

The difference in the path length between the direct and cascade \gr s leads to a
significant time delay of the cascade emission signal. For a given jet misalignment 
angle $\theta_{\rm obs}$, the time delay of 
emission coming from the direction $\theta$  away from the source is
$T_{\rm delay}\sim\frac{D}{c}\left(\frac{\sin\theta+\sin(\theta_{\rm obs}+\Theta_{\rm jet})}{\sin(\theta+\theta_{\rm obs}+\Theta_{\rm jet})}-1\right)\simeq\frac{D\theta(\theta_{\rm obs}+\Theta_{\rm jet})}{2c}\simeq
3\times 10^6\left[\frac{(\theta_{\rm obs}+\Theta_{\rm jet})}{5^\circ}\right]\left[\frac{\theta}{5^\circ}\right]
\mbox{ yr }$.
Comparing this time scale with the typical time scale of AGN activity, $T_{\rm AGN}\sim 10^7$~yr, one sees that degree-scale extended emission in the GeV energy range depends on the TeV \gr\ luminosity of the blazar integrated over its life-time.

\textit{Results of numerical modeling.}
To model the asymmetric  extended emission from the \gr\ initiated electromagnetic 
cascade in intergalactic space, we have extended our two Monte-Carlo codes such 
that they follow now the 
three-dimensional trajectories of individual cascade particles moving through the 
EGMF. The turbulent
component of the EGMF has been calculated following the algorithm of \citet{jokipii94}. 

To produce an image of the \gr\ induced electromagnetic cascade, as it would be 
detected by a \gr\ telescope, we use the algorithm described by \citet{elyiv09}. 
We have verified that the results obtained using the two different codes are 
compatible with each other.

We record positions and directions of all secondary \gr s which cross a sphere of 
the radius $R=D$ around the source. We choose the directions of primary \gr s to 
be distributed within a cone with an opening angle $\Theta_{\rm jet}$. We consider 
a primary \gr\ beam with a Gaussian profile, so that the probability for a primary 
\gr\ to have a direction misaligned by an angle $\Theta$ with respect to the jet axis 
is $p(\Theta)\sim \exp\left(-\Theta^2/\Theta_{\rm jet}^2\right)$.

For simplicity, we consider a monochromatic primary \gr\ beam with all the primary \gr s 
having the same energy $E_{\gamma_0}=1$~TeV. This is sufficient to demonstrate the
existence of the effect discussed here for the first time, namely degree-scale 
jet-like extensions in {\it Fermi\/} 
images of TeV blazars. The EBL background is taken from the calculations of 
\citet{kneiske04}. We fix the distance to the source as $D=400$~Mpc, so that 
$\tau(E_{\gamma_0})\sim 1$. The EGMF is chosen to have a correlation length of the order 
of several Mpc, with its power spectrum sharply peaked at the wavenumber 
$k\simeq 1$~Mpc$^{-1}$. Our results could be generalized in a straightforward way to 
the case of an arbitrary primary \gr\ spectrum, arbitrary distance to the source and 
different EBL models, when considering extended emission from particular TeV blazars 
with known TeV band spectra and known redshift. 

\begin{figure}
\includegraphics[width=\linewidth]{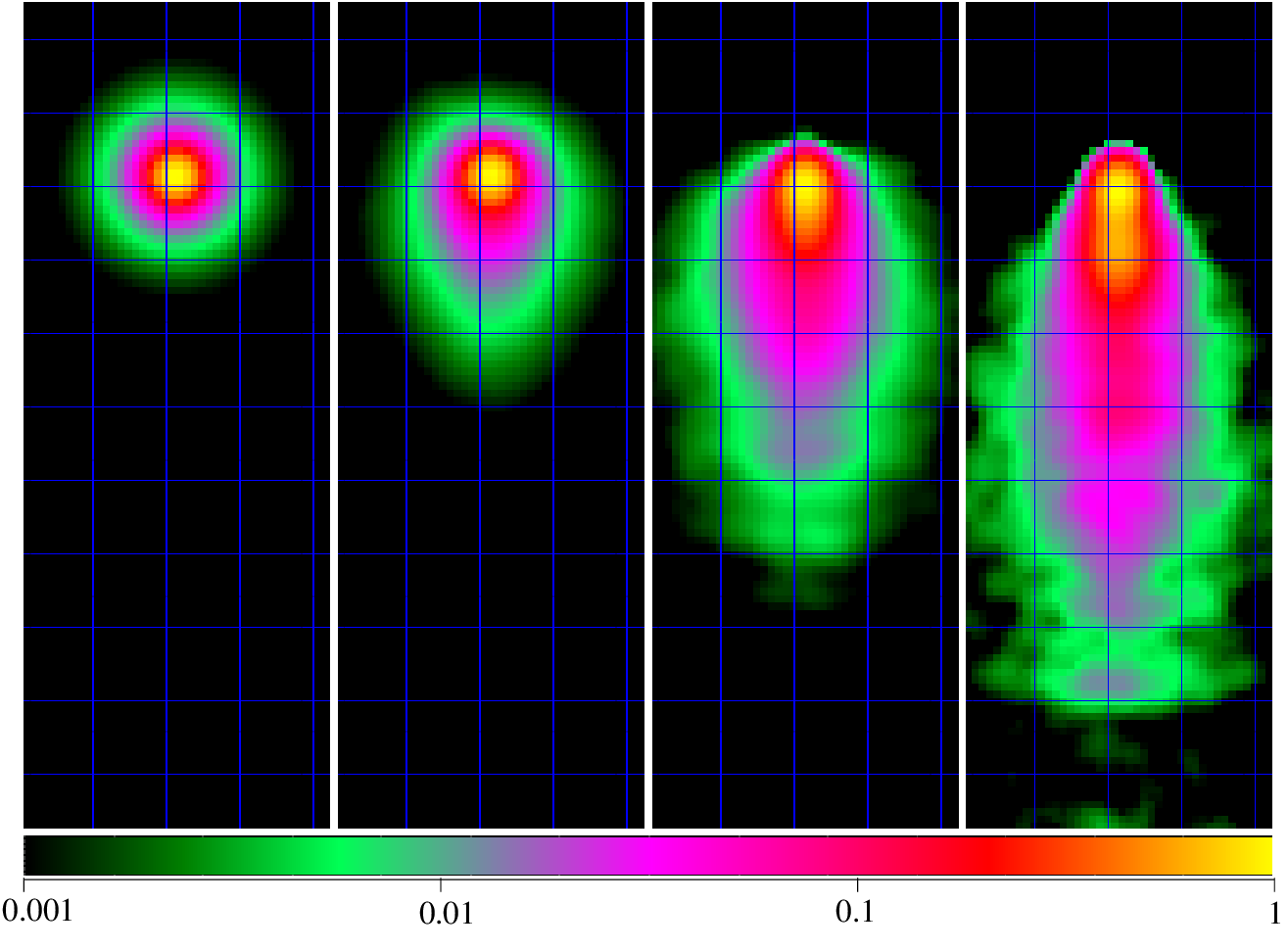}
\caption{$E>1$~GeV band images of the sky region around TeV blazars with jets inclined at $\theta_{\rm obs}=0^\circ$, $\theta_{\rm obs}=3^\circ$,  $\theta_{\rm obs}=6^\circ$ and $\theta_{\rm obs}=9^\circ$ (left to right). The jet opening angle is $\Theta_{\rm jet}=3^\circ$, the EGMF strength is $B=10^{-16}$~G.
The spacing of the coordinate grid is $2^\circ$, the color scale is logarithmic in surface brightness:
Yellow corresponds to the maximal surface brightness, black corresponds to the surface brightness less than $10^{-3}$ of the maximal value.}
\label{fig:offaxis}
\end{figure}
 
 Figure \ref{fig:offaxis} shows the effect of the misalignment of the primary \gr\ beam 
with the LOS on the morphology of the extended emission. The left panel of the Figure 
corresponds to the situation $\theta_{\rm obs}=0$, which is equivalent to the 
axially-symmetric case considered by \cite{kachelriess09} and \cite{elyiv09}. 
An axially-symmetric extended "halo" around the primary point source is clearly
visible. The other panels of the Figure show the cases of a jet with opening angle 
$\Theta_{\rm jet}=3^\circ$ misaligned by the angles $\theta_{\rm obs}=3^\circ$, $6^\circ$ and 
$9^\circ$, respectively. It is clear that the misalignment of the jet axis with the 
line of sight leads to the appearance of an extended jet-like feature on one side of 
the source. The ratio of the point source flux to the flux of the extension grows with 
the increase of the misalignment angle $\theta_{\rm obs}$.
 
\begin{figure}
\includegraphics[width=\linewidth]{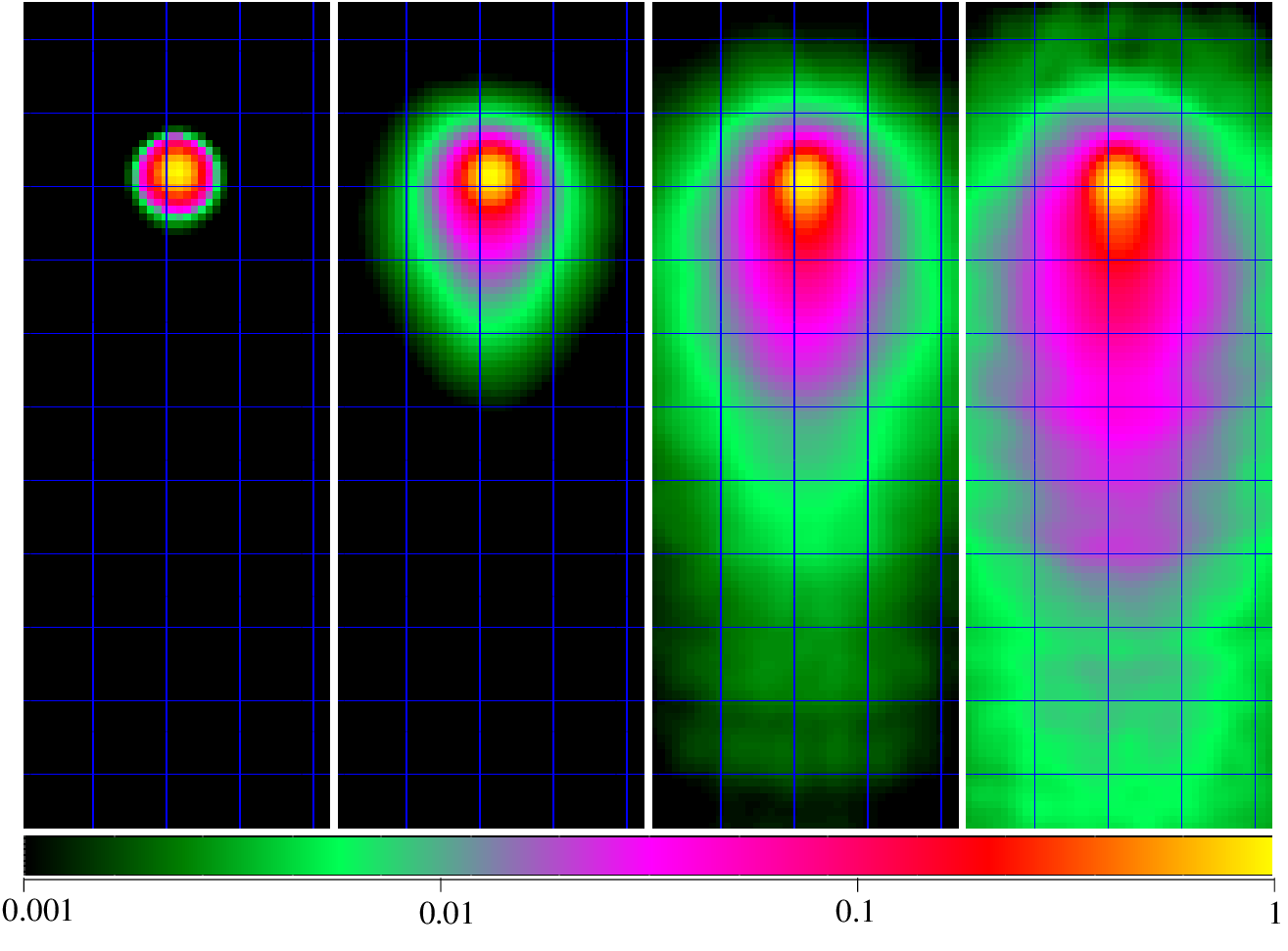}
\caption{$E>1$~GeV band images of the sky region around TeV blazars with $\Theta_{\rm jet}=\theta_{\rm obs}=3^\circ$  for different values of the EGMF strength. From left to right: $10^{-17}$~G, $10^{-16}$~G, $10^{-15}$~G, $10^{-14}$~G.  }
\label{fig:B}
\end{figure}

The angular extension of the cascade emission depends on the strength of the EGMF as long as 
the trajectories of $e^+e^-$ pairs are not completely randomized. The morphological properties 
of the jet-like emission are practically independent from the properties of the EGMF, when the EGMF 
strength is such that the deflection angle $\delta\ge 2\pi$. Figure \ref{fig:B} shows the growth of the source extension with the increase of 
the EGMF strength. For magnetic fields stronger than $B\sim 10^{-15}$~G,  the size of the extended 
source reaches ten(s) of degrees. In this case, the extended source could significantly contribute 
to the diffuse \gr\ background. 

\begin{figure}
\includegraphics[width=\linewidth]{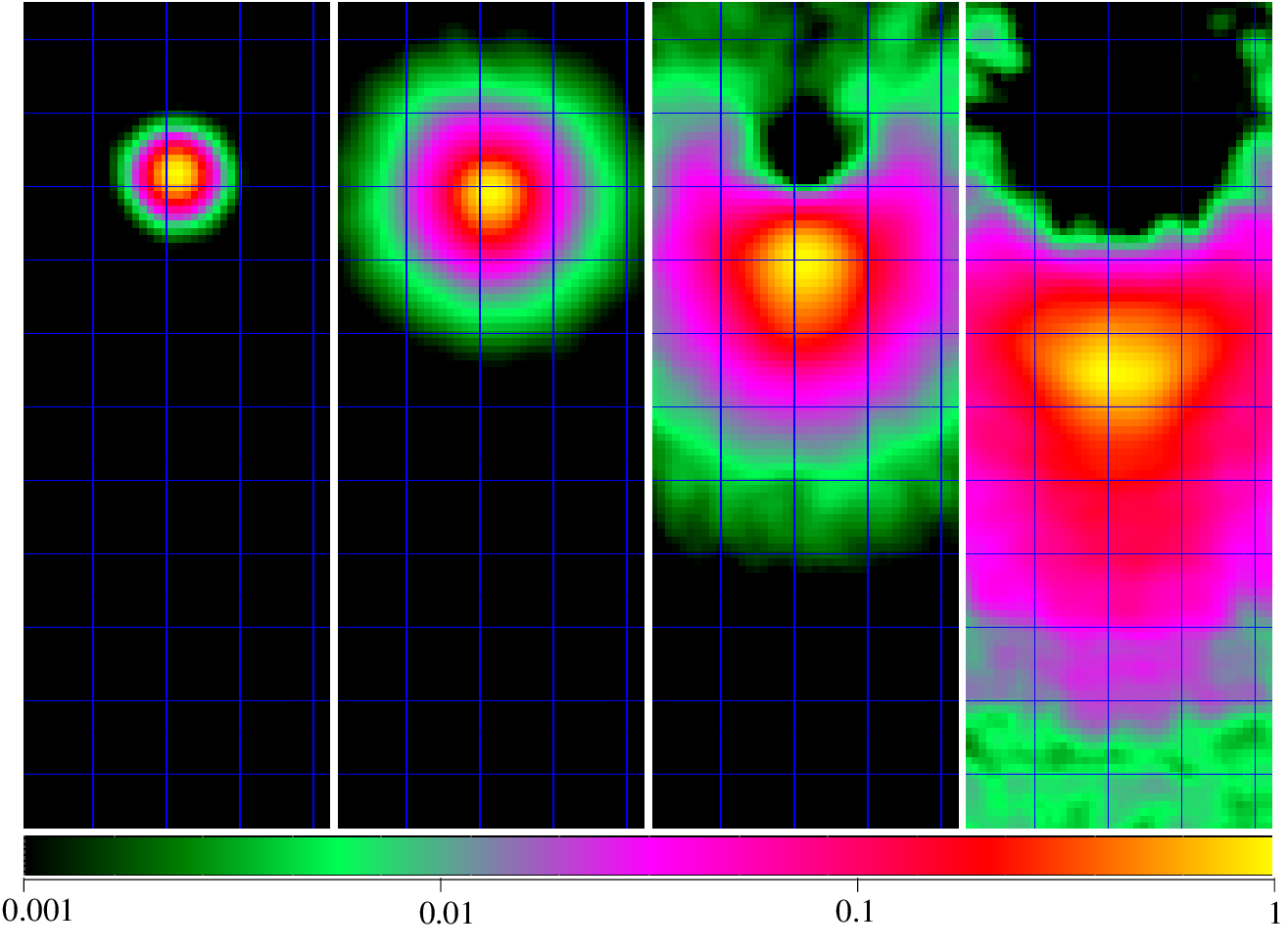}
\caption{$E>1$~GeV band images of the sky region around a TeV blazar with $\Theta_{\rm jet}=\theta_{\rm obs}=3^\circ$ at different times following instantaneous injection of 1~TeV \gr s at the source. From left to right: images in time intervals $0<T_{\rm delay}<10^5$~yr, $10^5$~yr$<T_{\rm delay}<10^6$~yr, $10^6$~yr$<T_{\rm delay}<3\times 10^6$~yr  and $3\times 10^6$~yr$<T_{\rm delay}<10^7$~yr after the outburst.  $B=10^{-16}$~G.}
\label{fig:time_sequence}
\end{figure}

Cascade emission coming from regions with angular distance $\theta\ge 1^\circ$ to the primary 
source is delayed by $T_{\rm delay}\sim 10^5$--$10^7$~yr compared to the direct emission from the 
source. 
This means that ``echos" from periods of enhanced activity 
of the source (e.g.\ an enhanced accretion rate following major merger episodes),  which 
happened all along the life-time of an AGN some time $T$ ago, could enhance the flux 
at the distance $\theta\simeq 1.7^\circ\left[T/10^6\mbox{ yr}\right]\left[(\theta_{\rm obs}+\Theta_{\rm jet})/5^\circ\right]$ from the source. 

Figure~\ref{fig:time_sequence} shows a time sequence of $E>1$ GeV band images of the sky 
region around a TeV source at different times after a short episode of TeV  \gr\ emission. 
One can clearly see that the emission at large angular distances is delayed by up to $10^7$~yr. 

The flux coming from the region at an angular distance $\theta$ from the point source is 
proportional 
to the source flux averaged over the period $T_{\rm delay}$. Therefore it is possible that 
GeV \gr s  are detectable today from an AGN which was active some $10^7$~yr ago, but is 
at present not active anymore. In this case a GeV source would be classified as ``unidentified": 
The parent AGN {\it (a)} could not be identified as an AGN in the optical, X-ray and TeV 
\gr\ bands or {\it (b)} the GeV source is displaced from the position of the parent AGN. 
The characteristic feature of such an unidentified ``AGN remnant" is the absence of 
counterparts at lower energies: If the GeV \gr s are produced by $e^+e^-$ pairs deposited in 
the intergalactic medium by primary TeV \gr s, the only energy loss mechanism for the pairs is 
IC scattering on CMB photons. 
 
\textit{Discussion.} The presence of extended jet-like emission at degree scales should be a 
generic feature of GeV band images of TeV blazars. The total flux of the jet-like extended source 
is proportional to the source luminosity in the TeV energy band. Taking into account the fact 
that TeV blazars have hard \gr\ spectra, the primary source luminosity in the TeV band could 
be much larger than its GeV luminosity, so that the overall extended source luminosity could 
be higher than the primary source luminosity in the GeV band. This means that the best candidates 
for the search of extended emission are TeV blazars with hard intrinsic spectra.

This does not automatically mean that the extended emission should be readily detectable in 
{\it Fermi\/} images of TeV blazars. In spite of the larger luminosity, the extended source 
flux might be suppressed if the EGMF is strong enough to randomize the trajectories of $e^+e^-$ 
pairs before they loose their energy to the GeV band via IC emission. The maximal 
possible suppression of the extended source flux is by a factor $\Theta_{\rm jet}^{-2}\sim 100$. 

Another potential problem for the detection of jet-like extended emission next to  TeV blazars 
is that the extended source has to be identified on top of the diffuse \gr\ background. 
The minimal detectable flux  for extended sources increases roughly as $\theta^{1/2}$, where $\theta$ is the angular length of the jet-like extended source. Thus sources at larger distances, for which 
the jet-like extensions appear more compact, are better candidates for the search of extended 
emission in the {\it Fermi\/} energy band.

Finally, the detectability of extended emission close to TeV blazars strongly depends on the 
angular resolution of the LAT telescope. At low energies, $E_\gamma\sim 0.1$~GeV, the LAT angular 
resolution is relatively poor, $\theta_{\rm PSF}\simeq 10^\circ$. It is clear that only very large 
angular size jet-like extensions with an angular diameter $\theta\sim\theta_{\rm PSF}$ could be 
detected. However, the detectability of such large extended sources would be complicated by 
the high level of diffuse sky background within the $\sim 10^\circ$ region around the source. 
At the same time, the size of the PSF decreases to $\theta_{\rm PSF}\le 1^\circ$ above GeV energies. 
This dramatically improves the sensitivity of the telescope for the search of extended emission: 
Extensions of much smaller angular size could be detected on top of a strongly reduced background. 
This favors the search of extended emission at energies above $\sim 1$~GeV. 

As an example, we consider a blazar with the TeV band luminosity $L_0 (E_{\gamma_0})\sim 10^{43}$~erg/s beamed in a cone with opening angle $\Theta_{\rm jet}=3^\circ$, so that the equivalent isotropic luminosity of the source is $L_{\rm iso}\simeq 10^{45}$~erg/s. In the absence of absorption, the source would give a flux $F_{\rm iso,0}\simeq 10^{-10}\left[D/300\mbox{ Mpc}\right]^{-2}$~erg/(cm$^2$s). At energies $E_\gamma$ such that $\tau (E_{\gamma_0})\ge 1$ the overall luminosity of the cascade emission is comparable to the primary source luminosity at the energy $E_{\gamma_0}$, so that in the case of small EGMFs ($\delta\le \Theta_{\rm jet}$), i.e.\ at the level of the lower bounds $B\sim 10^{-17}$ -- $10^{-16}$~G derived from {\it Fermi} observations \citep{neronov_vovk,tavecchio10}, the flux in the cascade is $F_{\rm cascade}\sim F_{\rm iso,0}$. In this case the cascade emission is readily detectable in the GeV band by {\it Fermi}. In the opposite case, the cascade emission is completely  isotropized  and the cascade flux is suppressed by a factor of $1/\Theta_{\rm jet}^2\sim 4\times 10^2$, so that it is marginally below the minimal detectable flux for extragalactic  {\it Fermi} point sources in the GeV band, $F_{\rm min}\sim 10^{-12}$~erg/cm$^2$s. 

Thus in the most pessimistic  case jet-like extensions are detectable only for the brightest blazars with the observed steady state flux at the level of $F_{\rm iso}\simeq \exp(-\tau)F_{\rm iso,0}\sim 3\times 10^{-11}$~erg/cm$^2$s at (assuming $\tau \simeq 1$). 
Only several extragalactic sources with sufficient steady state flux are detected by {\it Fermi} above 100~GeV \citep{neronov_IC310}. Among these sources, 3C 66A, 1ES 0502+675, PG 1553+113 and, possibly, PKS 2155-304 are at sufficiently large redshifts for strong cascade emission in the GeV energy range. These sources should be considered as primary candidates for the search of the jet-like extended emission. By contrast,
all sources with sufficiently high 
flux at few$\,\times 100$\,GeV
\citep{neronov_100GeV} are viable candidates for the detection of the  extended jet-like emission, if the EGMF is weak. 

The number of detectable "blazar remnants" also strongly depends on the EGMF. If the EGMF is so weak that $T_{\rm delay}\ll T_{\rm AGN}$, blazar remnants would 
not exist, since cascade and direct emission would be observed together.
If typical deflection angles of electrons emitting in the GeV band are several degrees ($B\sim 10^{-17}$~G for large $\lambda_B$), the number of blazar remnants observable in the GeV band should be comparable to the number of active TeV blazars, since  $T_{\rm delay}\sim T_{\rm AGN}$. If the EGMF is much stronger, the number of potentially observable blazar remnants grows because the cascade emission is emitted in a wider cone than emission from the parent blazar. However, the typical flux of the blazar remnants decreases because of the same effect. Thus, for strong EGMFs, the number of blazar remnants above the {\it Fermi} sensitivity limit might be very small. The strong dependence of the observability of blazar remnants on the EGMF strength implies that constraints on the EGMF could be deduced from their source statistics.


To summarize,
we have shown that GeV band images of TeV blazars should possess degree-scale jet-like extended 
features. These features trace the direction of the TeV \gr\ beam emitted by the blazar. 
They are produced as results of electromagnetic cascades initiated by TeV \gr s interacting 
with EBL photons. We have performed Monte Carlo simulations of three-dimensional electromagnetic 
cascades developing in the EGMF. Using these Monte Carlo simulations,  we have derived the 
properties of the GeV jet-like extended emission near TeV blazars. We have investigated the 
dependence of the characteristics of the jet-like extended sources (the angular size, the brightness 
profile) on the strength of the extragalactic magnetic fields and on the opening angle and 
orientation of the primary TeV \gr\ beam from the blazar. We have also demonstrated that 
the \gr\ signal in the jet-like extended emission is delayed up to $10^7$~years compared to 
the direct \gr\ signal from the primary point source. The very long time delay of the cascade 
emission means that the extended GeV source could be detected next to a blazar which is no 
longer active as blazar. 

\textit{Acknowledgments.} 
A.E.\ is supported by a fellowship from the Belgium Federal Science
Policy Office, A.N.\ by the Swiss National Science Foundation project 
PP00P2\_123426/1, and S.O.\  by a Marie Curie IEF fellowship from the 
European Community and by the Romforskning program of Nork Forskningsradet.

\end{document}